\documentclass{aastex62}

\graphicspath{{./}{figures/}}

\received{February 28, 2020}
\revised{\today}
\accepted{\today}
\submitjournal{ApJ}
\usepackage{chngcntr}
\counterwithout{figure}{section}
\usepackage{amsmath}
\usepackage{tikz}
\usepackage{float}
\usepackage{dcolumn}
\usepackage{bm}

\shorttitle{FRB Constraints on Dark Matter}
\shortauthors{Sammons et al.}
\begin{document}

\title{First Constraints on Compact Dark Matter from Fast Radio Burst Microstructure}

\correspondingauthor{Mawson W. Sammons}
\email{mawson.sammons@graduate.curtin.edu.au}

\author[0000-0002-4623-5329]{Mawson W. Sammons}
\affil{Curtin University, Perth, WA 6845, Australia}
\author[0000-0001-6763-8234]{Jean-Pierre Macquart}
\affiliation{International Centre for Radio Astronomy Research, Curtin Institute of Radio Astronomy, Curtin University, Perth, WA 6845, Australia}
\author{Ron D. Ekers}
\affiliation{CSIRO Astronomy and Space Science, PO Box 76, Epping, NSW 1710, Australia}
\affiliation{International Centre for Radio Astronomy Research, Curtin Institute of Radio Astronomy, Curtin University, Perth, WA 6845, Australia}
\author[0000-0002-7285-6348]{Ryan~M.~Shannon}
\affiliation{Centre for Astrophysics and Supercomputing, Swinburne University of Technology, Hawthorn VIC 3122, Australia}
\author{Hyerin Cho}
\affiliation{School of Physics and Chemistry, Gwangju Institute of Science and Technology, Gwangju, 61005, Korea}
\author{J. Xavier Prochaska}
\affiliation{University of California, Santa Cruz, 1156 High St., Santa Cruz, CA 95064, USA}
\author{Adam T. Deller}
\affiliation{Centre for Astrophysics and Supercomputing, Swinburne University of Technology, Hawthorn VIC 3122, Australia}
\author{Cherie K. Day}
\affiliation{Centre for Astrophysics and Supercomputing, Swinburne University of Technology, Hawthorn VIC 3122, Australia}
\affiliation{CSIRO Astronomy and Space Science, PO Box 76, Epping, NSW 1710, Australia}
\begin{abstract}
Despite existing constraints, it remains possible that up to $35\%$ of all dark matter is comprised of compact objects, such as the black holes in the 10-100\,M$_\odot$ range whose existence has been confirmed by LIGO. The strong gravitational lensing of transients such as FRBs and GRBs has been suggested as a more sensitive probe for compact dark matter than intensity fluctuations observed in microlensing experiments. Recently ASKAP has reported burst substructure down to $15\mu$s timescales in FRBs in the redshift range $0.3-0.5$. We investigate here the implications of this for the detectability of compact dark matter by FRBs. We find that a sample size of $\sim130$ FRBs would be required to constrain compact dark matter to less than the existing 35$\%$ limit with 95$\%$ confidence, if it were distributed along $\gtrsim 1\,$Gpc-long FRB sightlines through the cosmic web. Conversely, existing constraints on the fraction of compact dark matter permit as many as 1 in {\bf $\approx 40$} of all $z \lesssim 0.4$ FRBs to exhibit micro-lensed burst structure. Approximately $170$ FRBs intercepting halos within $\sim 50\,$kpc would be required to exclude the fraction of compact dark matter in each intercepted halo to a similar level. Furthermore, we consider the cumulative effects of lensing of the FRB signal by a macroscopic dark matter distribution. We conclude that lensing from a uniform distribution of compact objects is likely not observable, but suggest that FRBs may set meaningful limits on power-law distributions of dark matter.
\end{abstract}

\section{Introduction}
Dark matter comprises 24$\%$ of the energy density of the Universe \citep{bennett_nine-year_2013}, yet its indeterminate form represents one of the largest unsolved problems in astrophysics. Exotic particles from outside the standard model, such as Weakly Interacting Massive Particles (WIMPs) or axions have been invoked as possible explanations (see \cite{bertone_particle_2005} for a review). However, some fraction of dark matter could reside in the Universe as compact objects, such as black holes or neutron stars.

Decades of extensive research has constrained the fraction of dark matter present in compact objects over a range of masses. Low mass objects ($10^{-7}M_\odot\lesssim M\lesssim 10$M$_\odot$) are excluded as the dominant form of dark matter in the Milky Way and environs based on the absence of stellar variability caused by gravitational microlensing \citep{tisserand_limits_2007, wyrzykowski_ogle_2011, alcock_macho_1997}. High mass objects ($\gtrsim 100$M$_\odot$) are excluded by the lack of expected kinematic perturbations to wide binary orbits and ultra faint dwarf galaxies \citep{quinn_reported_2009, brandt_constraints_2016}. 

The only population of compact objects that are not well constrained lie in the range of 10 to 100M$_\odot$. There is a known population of black holes in this mass range; gravitational wave observations by LIGO have detected several mergers of these black holes \citep{ligo_scientific_collaboration_and_virgo_collaboration_gwtc-1:_2019}. Subsequent theories suggest that dark matter composed of $\sim30$M$_\odot$  primordial black holes could explain the merger event rates observed by LIGO \citep{bird_did_2016, clesse_clustering_2017, sasaki_primordial_2016}. Better constraints on the fraction of compact dark matter within the 10-100M$_\odot$ range could therefore be key in identifying some fraction of dark matter.

The strong gravitational lensing of extragalactic transients provides a way to either detect or to place more stringent constraints on dark matter. The strong lensing of type Ia supernovae has been used to limit the compact dark matter fraction to less than ~35$\%$ for all objects more massive than 0.01M$_\odot$ \citep{zumalacarregui_limits_2018}. Recently, it has been realised that cosmological transients such as Gamma Ray Bursts (GRBs) and Fast Radio Bursts (FRBs) will allow constraints to be placed at a much higher significance \citep{ji_strong_2018, laha_lensing_2018}. 

In both cases strong lensing creates multiple images of the source. Unlike the gravitational lensing of quasars by foreground galaxies \citep{wong_h0licow_2019}, the images formed by a compact object would be too close to be spatially resolved. However, the images of the source will arrive separated in time due to different gravitational and geometric time delays along each path. This temporal separation ($\Delta t$) is linearly dependent upon the lens mass, whereas the magnification ratio is mass independent. The lens mass and geometry can be constrained using these two observables. Due to the achromatic nature of gravitational lensing the same formalism initially suggested by \cite{munoz_lensing_2016} for FRBs can be applied at all wavelengths. The formalism ignores the effect of physical optics, which becomes important when the Einstein radius of the lens is smaller than the Fresnel scale, $\sim \sqrt{D_{\rm eff} \lambda/2 \pi}$, where $D_{\rm eff}$ is the effective distance to the lens.  This occurs for lens masses less than $\sim 10^{-5}$M$_\odot$ at a frequency of 1\,GHz, and is well below the masses considered here; hence a full wave optics treatment, such as that explored by \cite{jow_wave_2020}, is not yet warranted.

Several thousands of GRBs have been discovered at redshifts $z\sim1$ by dedicated GRB observatories such as \textit{Swift}, \textit{BATSE}, and \textit{Fermi}. The cosmological distances they traverse allow them to probe a large volume of the Universe for compact dark matter. GRBs have a broad temporal profile ranging from milliseconds to minutes \citep{ji_strong_2018}, and as a result, distinguishing multiple images is more difficult as the time delay between signals lensed by a 10M$_\odot$ compact object will be less than the duration of the GRB. \cite{ji_strong_2018} have proposed auto-correlating the light curve as a method of detecting lensing. They conclude, however, that current GRB observatories would need to reduce their noise power by at least an order of magnitude to be able to detect lensing in the 10-100M$_\odot$ mass range.

In contrast, FRBs have temporal profiles ranging from tens of microseconds \citep{cho_spectropolarimetric_2020} to several milliseconds, which is often shorter than the anticipated delay ($\sim 1$ms) for lensing by compact objects in the mass range under consideration here. This enables multiple images to be clearly distinguished, hence rendering FRBs considerably cleaner probes of compact structure along their sightlines. FRBs are highly luminous, extragalactic radio pulses, and those such as FRB\,181112 \citep{cho_spectropolarimetric_2020} with substructure on timescales of a few tens of microseconds provide, to date, the finest timescale probe of sightlines at cosmological distances.  Moreover, a unique capability of radio interferometric observations of such bursts is their ability to directly capture the {\it wavefield} of the each FRB at extremely high time resolution \citep[3 ns; see][]{cho_spectropolarimetric_2020}. This affords a powerful new diagnostic of the presence of gravitational lensing.  The wavefield, which is directly observable at radio wavelengths, of any pair of paths in the lensed signal should be correlated, whereas burst substructure intrinsic to the FRB would not.

Of the FRBs localised to host galaxies so far, all have been at redshifts $z<1$, placing the current sample generally closer in the Universe than GRBs \citep{ji_strong_2018, coward_swift_2013, bannister_single_2019, prochaska_low_2019}.  However, this limitation can be overcome by inferring source redshifts from the dispersion measures of non-localised FRBs \citep{macquart_census_2020}; the existence of FRBs with dispersion measures exceeding $2000\,$pc\,cm$^{-3}$ \citep[e.g.][]{bhandari_survey_2018} ostensibly places some fraction of the population at $z>2$.

To detect strong lensing, the temporal separation must be sufficiently large to allow each image to be distinguished. This is constrained by the shortest distinct temporal structure in the signal. In this paper we examine the implications of the high-time resolution structure observed in the FRBs 180924 and 181112.  In FRB\,180924 the shortest timescale corresponds to its rise time of only $30\mu$s \citep{farah_private_2020}. FRB\,181112 has recognisable temporal structure on the scale of $15\mu$s, the shortest structure observed in an extragalactic radio signal \citep{cho_spectropolarimetric_2020}. The resolution of temporal structures of $\sim 10\mu$s enables searches for lensing at temporal separations an order of magnitude below those considered in previous treatments \citep[0.1\,ms; ][]{munoz_lensing_2016, laha_lensing_2018}. The S/N of recorded FRBs allows us to consider magnification ratios an order of magnitude above previous treatments \citep[$<$5; ][]{munoz_lensing_2016, laha_lensing_2018}. Additionally, if the FRB passes close to an intervening galaxy, as it did for FRB\,181112 \citep{prochaska_low_2019}, it opens up the possibility  of examining lensing attributable to a specific galaxy along the burst sightline, other than the host galaxy or the Milky Way. \cite{munoz_lensing_2016} and \cite{laha_lensing_2018} report that a sample of 10$^4$ FRBs would be required to exclude the compact dark matter fraction to less than 1$\%$.  
Assuming a $\Lambda$CDM cosmology, we apply the same formalism to estimate the constraining potential of detected high time resolution FRBs comparable to FRBs\,181112 and 180924.  
\section{\label{sec:Theory}Theory}
In the weak field limit, where the gravitational potential $|\Phi|\ll c^2$, gravitational lensing can be modelled as an achromatic deflection of incident light by a thin screen. Under this treatment, a point mass lens will produce two images on the lens plane. The temporal separation, magnification ratio and position of these images are determined by the angular impact parameter of the source ($\beta$) normalised by the characteristic Einstein radius of the lens ($y=\beta/\theta_E$).

Here we briefly review previous theory as expounded by \cite{munoz_lensing_2016} and \cite{laha_lensing_2018}. Following this formalism, the difference in arrival time between the images ($\Delta t$) and the ratio of each magnification ($R_f$) correspond to unique normalised angular impact parameters of the source $y_{\Delta t}$ and $y_{R_f}$, respectively. The relation between $y_{R_f}$ and $R_f$ can be expressed analytically as \citep{munoz_lensing_2016},
\begin{equation}\label{eq:yMag}
    y_{R_f}=\sqrt{\frac{R_f+1}{\sqrt{R_f}}-2},
\end{equation} 
which is notably independent of the lens mass. Conversely, $y_{\Delta t}$ cannot be derived analytically and is found numerically from \cite{munoz_lensing_2016}:
\begin{equation}\label{eq:timeDelay}
    \Delta t=\frac{4GM_L}{c^3}(1+z_L)\left[\frac{y}{2}\sqrt{y^2+4}+\ln\left(\frac{\sqrt{y^2+4}+y}{\sqrt{y^2+4}-y}\right)\right],
\end{equation}
where $M_L$ and $z_L$ are the mass and redshift of the lens, respectively.

To detect gravitational lensing, we require the normalised angular impact parameter to be within the observable range ($y_{\text{min}}$--$y_{\text{max}}$). This range is defined by two conditions: (1) The associated time delay calculated from eq. (\ref{eq:timeDelay}) must be less than the maximum observable time delay $\Delta t_{\text{max}}$ and greater than the minimum distinguishable separation $\Delta t_{\text{min}}$. The length of the observation sets $\Delta t_{\text{max}}$, and $\Delta t_\text{min}$ is set by the structure in the pulse profile \citep{munoz_lensing_2016}. (2) The magnification ratio must be below the maximum ($\bar{R_f}$) set by the detection threshold \citep{munoz_lensing_2016}. 

For the thin screen approximation to be valid, the gravitational field at the impact parameter must also satisfy the weak field condition:
\begin{equation}\label{eq:weakFieldLimit}
    y_{\text{min}}\gg\frac{R_S}{D_L\theta_E}
\end{equation}
where $R_S$ and $D_L$ are respectively, the Schwarzschild radius and angular diameter distance of the lens. 
$y_{\text{min}}$ and $y_{\text{max}}$ define the annulus of the cross section to observable lensing. This cross section can then be used to calculate the observable lensing optical depth. Details on this calculation are provided in the following subsections for different environments. If the fraction of all dark matter that is compact ($f_{\text{DM}}$) is assumed to be constant, the probability of observing lensing ($P_L$) at least once in a set of $N$ FRBs can then be calculated as 
\begin{equation}
    P_L=1-\text{exp}\left[-\sum\limits_i^N\tau_i\right].\label{eq:prob}
\end{equation}
where $\tau_i$ is the optical depth of the ith FRB in the set. To exclude compact dark matter fractions of $\geq f_{\text{DM}}$ with 95$\%$ confidence, we require a null observation of lensing in a set of FRBs with a cumulative observable lensing optical depth of 3.0.
\subsection{Lensing in galaxy halos}
If we assume that compact dark matter takes the form of MACHOs (MAssive Compact Halo Objects), the only contribution to the lensing optical depth will come from the intervening galactic halos. In the local potential of a galaxy, the Hubble flow can be ignored and the optical depth calculated simply as
\begin{align}
    \tau&=\frac{f_{\text{DM}}\Sigma_{\text{halo}}}{M_L}\sigma\nonumber\\
        &=\frac{4\pi G f_{\text{DM}}\Sigma_{\text{halo}} }{c^2}\frac{D_LD_{LS}}{D_S}\left[y_{\text{max}}^2-y_{\text{min}}^2\right]\label{eq:optDGal}
\end{align}
where $\Sigma_{\text{halo}}$ is the halo mass surface density; $D_L$, $D_{LS}$, and $D_S$ are the angular diameter distances from the observer to the lens, from the lens to the source and from the observer to the source respectively; $\sigma$ is the observable lensing cross section, as described by \cite{laha_lensing_2018}; and $M_L$ is the mass of the individual compact objects in the halo. We assume a Navarro-Frenk White (NFW) dark matter distribution \citep{navarro_structure_1996} for which $\Sigma_{\text{halo}}$ has been derived by \cite{bartelmann_arcs_1996}.
\subsection{Lensing in the Intergalactic Medium}
Stellar remnants unbound from their host galaxies via natal kicks or gravitational interactions present a possible source of lensing in the intergalactic medium (IGM)\citep{atri_potential_2019}, as do primordial black holes. Here, the effects of the Hubble flow cannot be ignored. As derived in \cite{munoz_lensing_2016} and \cite{laha_lensing_2018}, the optical depth to lensing of a single source by a single compact object in the IGM is
\begin{align}
    \tau&=\int\limits_0^{z_s}d\chi(z_L)(1+z_L)^2n_{\text{IGM}}\sigma\nonumber\\
        &=\frac{3}{2}f_{\text{DM}}\Omega_c\int\limits_0^{z_s}dz_L\frac{H_0^2}{cH(z_L)}
        \frac{D_LD_{LS}}{D_S}(1+z_L)^2[y_{\text{max}}^2-y_{\text{min}}^2]\label{eq:optDIGM},
\end{align}
where $\chi$ is the co-moving distance, $n_{\text{IGM}}$ is the average co-moving number density of the lens, $H(z_L)$ is Hubble's constant at the lens redshift and $\Omega_c$ is the current density of dark matter. 

Both the halo and IGM lensing optical depths are separated into magnification and time delay-limited domains over which $y_{\text{max}}$ is limited by the corresponding condition. At low masses, $y_{\text{min}}$ increases until $y_{\text{min}}=y_{\text{max}}$, and the optical depth to lensing becomes zero. The halo and IGM lensing optical depths are mass independent over a large range of lens masses. This can be understood by considering equations (\ref{eq:optDGal}) and (\ref{eq:optDIGM}), respectively. The product of the Einstein radius squared and the projected number density is mass independent. Hence, by expressing the cross section in terms of the normalised angular impact parameters $y_{\text{min}}$ and $y_{\text{max}}$, the source of the mass dependence in each optical depth becomes isolated to $y_{\text{min}}$ and $y_{\text{max}}$. In the magnification-limited domain, $y_{\text{max}}$ is given by $y_{R_f}$ and will be independent of the mass (eq. (\ref{eq:yMag})). If $y_{max}$ is also much greater than $y_{\text{min}}$, then the optical depth to observable lensing in either the halo or IGM case will be effectively mass independent. The domain of this mass independent regime is determined by the minimum and maximum temporal separations.

\section{Results}

\begin{table*}
\caption{\label{tab:FRBparams}Observational parameters for localised high time resolution FRBs}
\begin{ruledtabular}
\begin{tabular}{ccccc}
 FRB &$\Delta t_{\text{min}}$(s)\footnote{$\Delta t_{\text{min}}$ and $\Delta t_{\text{max}}$ are defined respectively as the minimum and maximum observable time delays}&$\Delta t_{\text{max}}$(s)&$\bar{R_f}$\footnote{$\bar{R_f}$ is defined as the maximum magnification ratio, set by the detection threshold}
&Source Redshift\\ \hline
 181112\footnote{Intercepted a foreground galaxy at z=0.3674}&$15\times10^{-6}$&1.369&73.3&0.47550\\
 180924&$30\times10^{-6}$&1.445&64.7&0.3214\\
\end{tabular}
\end{ruledtabular}
\end{table*}

The determination of the redshift of an FRB, either by localisation or by inference from its dispersion measure \citep{macquart_census_2020}, allows the formalism outlined in Section \ref{sec:Theory} to be applied. Here, we calculate the halo and IGM lensing optical depth for localised FRBs 181112 \citep{prochaska_low_2019} and 180924 \citep{bannister_single_2019}. The temporal microstructure of these bursts has been resolved, enabling us to probe to the minimum value of $y_{\text{min}}$ allowed by the burst structure. FRBs 181112 and 180924 probe a similar range of masses ($0.1$M$_\odot\lesssim $M$\lesssim 10^4$M$_\odot$) due to their similar minimum and maximum temporal separations (Table \ref{tab:FRBparams}). Over this range of masses, eq.\,(\ref{eq:weakFieldLimit}) is satisfied, and the strong field region is orders of magnitude smaller than the spatial scale probed by a temporal separation of $10\,\mu$s. This is the scale of the smallest distinguishable temporal separation amongst known FRBs; therefore, the weak field approximation is valid for all cases considered here. 

\begin{figure}
    \centering
    \includegraphics[width=0.8\textwidth]{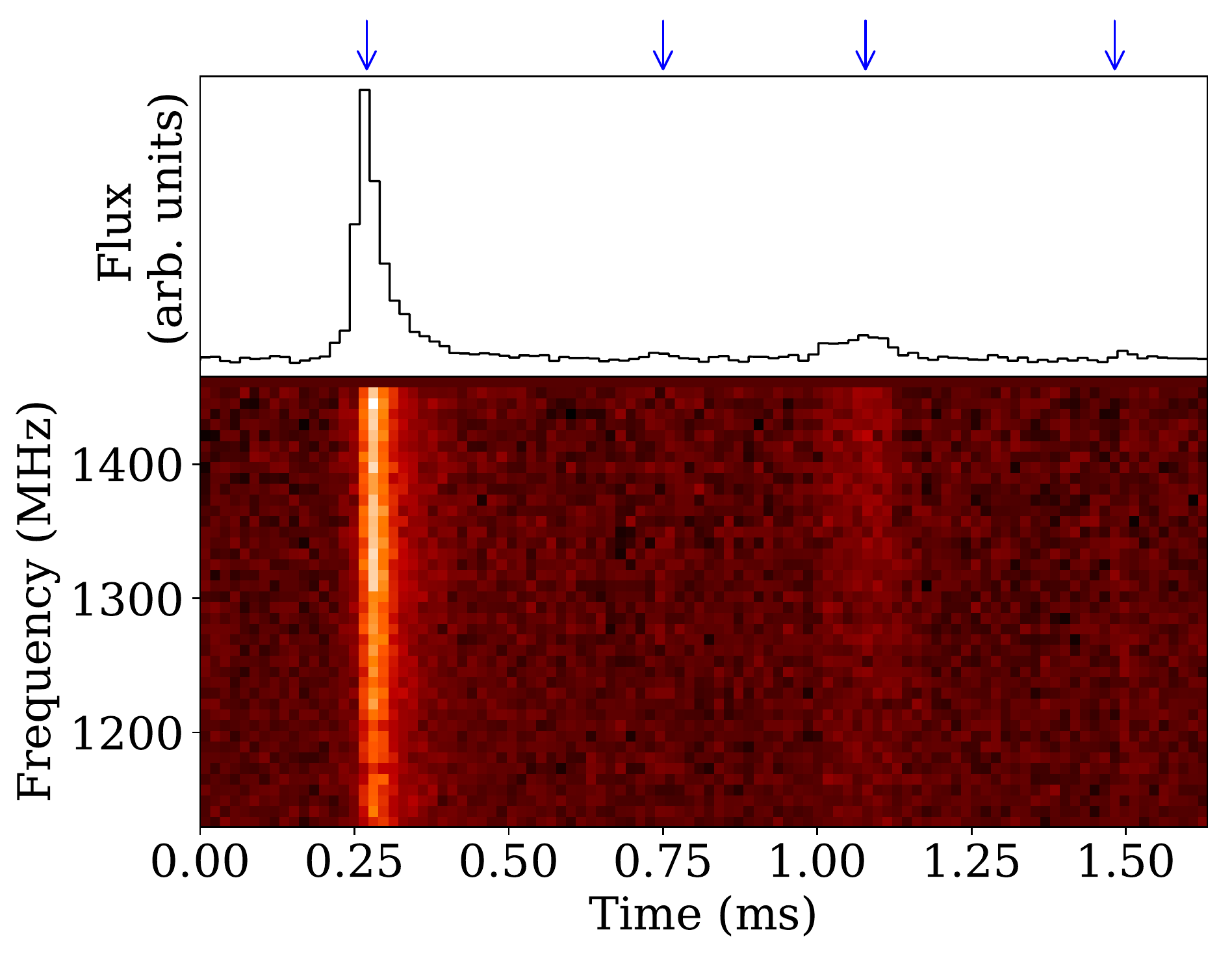}
    \caption{The pulse profile (top) and dynamic spectrum (bottom) of FRB\,181112 at 16$\mu$s and 8MHz temporal and spectral resolution respectively (This representation is smoothed to $16\mu$s to optimise the S/N, it is not the instrumental resolution). The pulse is seen to consist of two bright sub-pulses, at $t=0.25\,$ms and $1.1\,$ms, and two weaker sub-pulses at $t=0.75\,$ms and $1.50\,$ms, as indicated by the blue arrows.}
    \label{fig:181112TempProfile}
\end{figure}

The spectra of FRB\,181112 shown in Fig. \ref{fig:181112TempProfile} \citep[see also][]{cho_spectropolarimetric_2020} exhibits multi-peaked structure that could potentially be explained by gravitational lensing. Indeed, if the two major peaks are assumed to be two images, the temporal profile is consistent with gravitational lensing by a $\sim 10$M$_\odot$ compact object in the halo of the foreground galaxy (hence referred to as FG\,181112). \cite{cho_spectropolarimetric_2020} test for the presence of microlensing by searching for correlations in the burst wavefield with time; in the case of FRB\,181112 no fringes between sub-pulses were seen, suggesting the pulse multiplicity is more likely intrinsic to the burst, rather than multiple lensed copies of the same burst. However, the absence of a correlation is not definitive since other effects, notably due to differences in any turbulent cold plasma encountered along the slightly separated sightlines of the lensed images, could scatter the radiation in different manners, and thus destroy the phase coherence between the lensed signals.  However, in the present instance \cite{cho_spectropolarimetric_2020} also find that the polarization properties of the sub-bursts differ in detail, particularly in their circular polarization, an effect which is difficult to attribute to lensing \footnote{We can exclude circular polarization differences due to the existence of any relativistic plasma from a neutron star along the sightline, except at the source (where its presence would be irrelevant for the present argument).  The lens mass of $10$M$_\odot$ required to explain the sub-burst time delays is significantly above the largest observed neutron star mass of 2.14M$_\odot$ \citep{cromartie_relativistic_2020}, ruling out neutron stars as potential lens candidates, and the effects of any relativistic plasma associated with them.}.

As recorded in Table \ref{tab:FRBparams}, FRB\,181112 had an extremely narrow pulse profile, with its shortest temporal structure being 15 $\mu s$. FRB\,180924 had an extended scattering timescale of 580$\,\mu$s but a short rise-time of 30$\,\mu$s.  Were any delayed lensed signal present, it would also have a sharp 30$\,\mu$s rise time which would have been detectable within the tail of the overall pulse envelope. The maximum magnification ratios ($\bar{R_f}$) and redshifts are similar for each burst. To calculate $\bar{R_f}$, the S/N (signal-to-noise ratio) of the primary peak is divided by the detection threshold (3$\sigma$). A key difference between the two is that FRB\,181112 passed within 29kpc of FG\,181112, allowing it to probe a longer path through a galactic halo. In the following optical depth calculations these parameters are used to determine $y_{\text{min}}$ and $y_{\text{max}}$ from the equations defined in Section \ref{sec:Theory}. For all following calculations, we use values for $H_0$ and the cosmological density parameters from the Planck 2018 results \citep{planck_collaboration_planck_2018}.
\newpage
\subsection{Halo lensing optical depth}
The observable lensing cross section peaks approximately midway between the source and the observer and is minimal in both the host galaxy and the Milky way. Using the code of \cite{prochaska_frbsfrb_2019} we expect approximately one in 20 FRBs to intercept a foreground halo larger than $10^{12}$M$_\odot$ within $50\,$kpc. This is consistent with recent optical followups of arcsecond-localised FRBs, including FRB\,180924, which do not intercept massive galaxy halos within $\sim 50\,$kpc \citep{bannister_single_2019,chatterjee_direct_2017,marcote_repeating_2020}. Consequently, these FRBs are of negligible value in constraining the dark matter halos of specific galaxies. FRB\,181112, however, passes through a foreground galaxy where the cross section to lensing is much greater, making it an ideal candidate to constrain halo lensing. 

\begin{figure}
    \centering
    \includegraphics[width=0.8\textwidth]{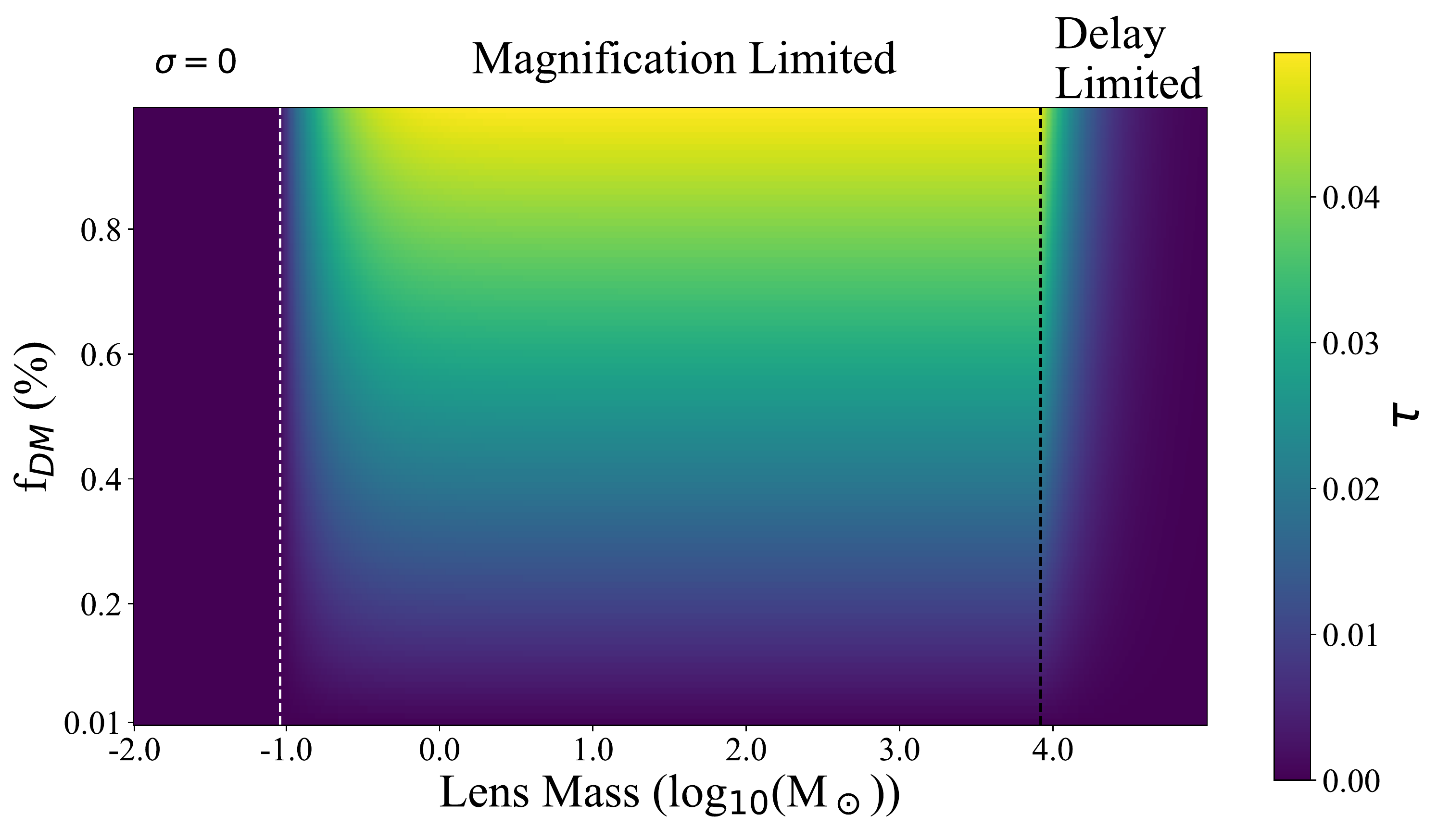}
    \caption{Optical depth to observable strong gravitational lensing by a point mass compact object of mass $M_L$ probed by FRB\,181112. For masses below the black dotted line $y_{\text{max}}$ is limited by the maximum magnification ratio, above $y_{\text{max}}$ is limited by the maximum time delay. The white dotted line marks the mass where $y_{\text{min}}$=$y_{\text{max}}$ and $\sigma=0$. This calculation assumes an NFW distribution of compact objects of a single mass $M_L$, comprising a fraction $f_{\text{DM}}$ of the host, Milky Way and foreground galaxies. FG\,181112 is modelled to have a halo virial mass $M_{\text{halo}}=10^{12}$M$_\odot$\citep{prochaska_low_2019} and a concentration parameter of $c\sim 7$.}
    \label{fig:galOD}
\end{figure}

Fig. \ref{fig:galOD} displays the optical depth to observable lensing by MACHOs probed by FRB\,181112. This optical depth is dominated by the contribution from the halo of FG\,181112. FG\,181112 is classified as a Seyfert galaxy with an old $10^{10.69}$M$_\odot$ stellar population \citep{prochaska_low_2019}. 

The white dotted line in Fig. \ref{fig:galOD} marks where the cross section to lensing becomes zero ($y_{\text{min}}=y_{\text{max}}$). Between this cutoff and a lens mass of $\sim1$M$_\odot$, $y_{\text{min}}$ and $y_{\text{max}}$ are comparable, and the optical depth to observing lensing depends on the mass of the lens. Above a lens mass of $\sim1$M$_\odot$, $y_{\text{max}}\gg y_{\text{min}}$ and the optical depth in the magnification-limited domain is approximately independent of mass. In the time delay-limited domain, the optical depth decreases sharply as a function of mass. We estimate that to conclude with 95$\%$ confidence that the MACHO dark matter fraction is less than 35$\%$, we require $\sim 170$ FRBs that intersect a foreground galaxy similar to FRB\,181112. This estimate is projected from the optical depth $\tau\approx0.018$ probed by FRB\,181112 at $f_{DM}=0.35$. Additionally, we can conclude with 90$\%$ confidence that the total mass, in the halo of the FG galaxy of FRB\,181112, contained in uniformly distributed compact objects may be no more than $4.5\times10^{13}$M$_\odot$.

\subsection{Lensing by structure in the cosmic web}
\begin{figure}
    \centering
    \includegraphics[width=0.8\textwidth]{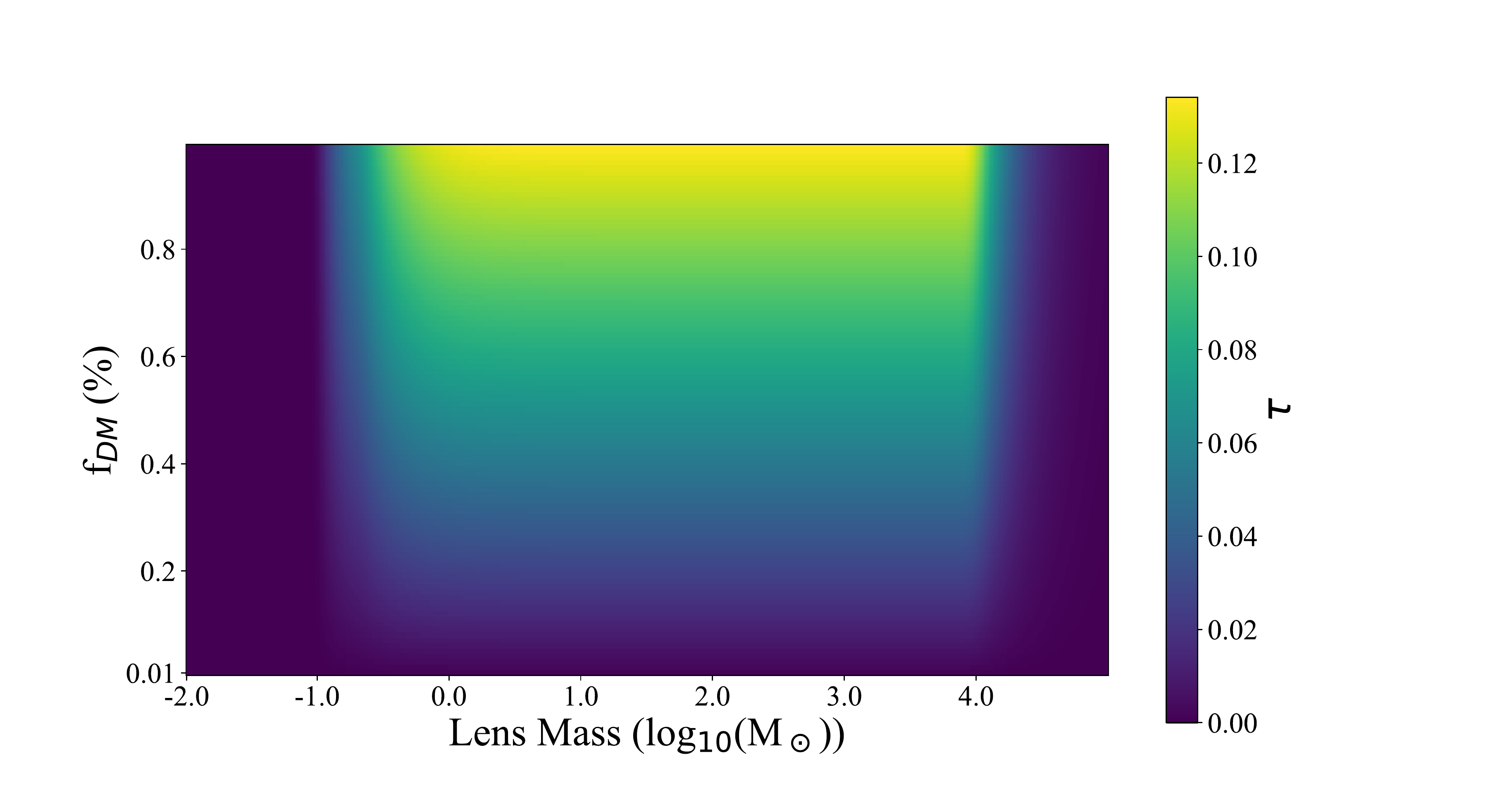}
    \caption{Cumulative optical Depth to observable strong gravitational lensing  by a point mass compact object of mass $M_L$ in the IGM probed by FRB\,181112 and FRB\,180924. We assume a uniform distribution of $M_L$ mass compact objects in co-moving space comprising a fraction $f_{\text{DM}}$ of the total dark matter of the Universe.}
    \label{fig:IGMOD}
\end{figure}

Fig. \ref{fig:IGMOD} displays the optical depth to lensing by a compact object due to any compact dark matter present throughout the cosmic web by FRB\,181112 and FRB\,180924, assuming that the dark matter density along their sightlines are representative of the mean cosmological dark matter density $\Omega_{\rm DM}$. This case shows the same trends as the halo lensing case, albeit with a much higher overall optical depth. Unlike in the halo lensing case, compact objects can be encountered anywhere in the path of an FRB. As a result, FRBs probe a much greater optical depth to lensing in this scenario. We estimate that to exclude compact dark matter fractions above $35\%$ with $95\%$ confidence, a comparatively smaller sample of $\sim 130$ FRBs would be required. Furthermore, this sample may be comprised of any observed FRBs. This estimate is projected from the average optical depth $\tau\approx0.024$ probed by FRB\,181112 or FRB\,180924 at $f_{DM}=0.35$. Under different an assumed FRB redshift distribution, \cite{laha_lensing_2018} and \cite{munoz_lensing_2016} estimate that to exclude $f_{DM}\geq 1\%$ with 99\% confidence, $10^4$ FRBs would be required.

\subsection{Gravitational Scattering} 
So far our treatment has been restricted to lensing by a single point mass. However, it is possible in principle that an ensemble of low mass clumps could collectively lens an FRB signal, characterising it with an achromatic, exponential scattering tail \citep{macquart_scattering_2004}.

We are thus motivated to examine whether gravitational scattering, caused by a cloud of substructure within a dark matter halo is observable. Within FRB\,181112, we do not observe a clear exponentially decaying scattering tail, placing an upper limit to the scattering timescale $\sim20\mu$s\citep{cho_spectropolarimetric_2020}. The lack of this feature was interpreted as a lack of turbulent plasma along the line of sight, as discussed in \cite{prochaska_low_2019}. However, it also places a constraint on the mass of lensing substructure in the intervening halo. Here we introduce the relevant theory and present some cursory constraints, leaving a more exhaustive treatment to a future paper.
In the limit where a large number of lenses exist within the coherence area, a statistical approach is mandated, and the characteristic delay timescale is, analogous to scattering in an inhomogeneous plasma \citep{macquart_scattering_2004},
\begin{equation}\label{eq:scatteringTime}
    t_{\text{scatt}}=\frac{1}{2\pi\nu}\frac{r_F^2}{r_{\text{diff}}^2},
\end{equation}
where $r_F$ is the Fresnel radius given by 
\begin{equation}\label{eq:fresnel}
    r_F^2=\frac{cD_LD_{LS}}{2\pi\nu D_S(1+z_L)}
\end{equation}
and $r_{\text{diff}}$ is the length scale over which the mass density fluctuations cause the gravitational phase delay to fluctuate by one radian RMS. 

To solve for the diffractive scale we must consider the RMS phase difference in the fluctuations over varying scales of the mass distribution. This quantity can be calculated from phase structure function \citep{macquart_scattering_2004}
\begin{equation}\label{eq:PSF}
    D_\psi(r)=\langle[\psi(\Vec{r'}-\Vec{r})-\psi(\Vec{r'})]^2\rangle=2K^2\int d^2\Vec{q}\left[1-e^{i\Vec{q}\cdot\Vec{r}}\right]q^{-4}\Phi_\Sigma (\Vec{q}),
\end{equation}
which is the means square difference in phase fluctuations as a function of the separation ($\Vec{r}$), spatial wavevector ($\Vec{q}$) and the mass surface density power spectrum ($\Phi_\Sigma (\Vec{q})$), in keeping with our thin screen approximation. 

In a simple model, where we assume a Poisson distribution of clumps (i.e. the number of clumps in any given area will be sampled from a Poisson distribution with an average density of $\Sigma$), eq. (\ref{eq:PSF}) gives \citep[see, e.g.,][]{macquart_scattering_2004}
\begin{equation}\label{eq:rdiffUniform}
    r_{\text{diff}}=2.2\times10^2(1+z_L)^{-1}\nu^{-1}\left(\frac{M}{1\,{\rm M}_\odot}\right)^{-1}\left(\frac{\Sigma}{100\text{ clusters pc}^{-2}}\right)^{-1/2} \text{pc},
\end{equation}
where $\Sigma$ is the projected number density of clusters, assumed to be locally uniform. To achieve a smooth scattering tail, the number of lenses must exceed unity within the coherence area $\sim \pi r_F^2$, to the point where the discrete contributions of individual lenses would be indiscernible. Assuming that the halo of FG\,181112 obeys a NFW profile with a scale radius $R_s=24\,\text{kpc}$ and a virial mass $10^{12}$\,M$_\odot$\citep{prochaska_low_2019}, this would require a lens mass $M_L\ll\Sigma\pi r_F^2\approx3.7\times10^{-8}$M$_\odot$ for a Fresnel scale of $\sim 3$AU and an impact parameter of 29\,kpc (even if all the matter were contained in clumps of this size). The characteristic time delay for gravitational scattering at 1.2\,GHz would therefore be much less than $\sim3.3\times10^{-14}$s (from eq. (\ref{eq:scatteringTime})). Thus, we do not expect to observe any scattering tail associated with the lensing from a distribution of compact objects with a uniform density.

Under a CDM/WDM treatment, it is plausible that galactic dark matter could cluster following a spatial power law \citep{macquart_scattering_2004}, similarly to turbulent distributions of neutral gas and ionised plasma, which have been observed to have spectral indexes of $3\lesssim \beta \lesssim 4$ \citep{armstrong_electron_1995, dickey_southern_2001, stanimirovic_velocity_2001}.
\citep{macquart_scattering_2004}. A power law spectrum of mass density fluctuations can be projected onto a screen of thickness $\Delta L$ to give the mass surface density power spectrum required in eq. (\ref{eq:PSF}). Following the derivation of the phase structure function in \cite{macquart_scattering_2004}, for a power spectrum with an index $\beta$, between some inner ($l_0=1/q_\text{max}$) and outer ($L_0=1/q_\text{min}$) scales gives:
\begin{equation}\label{eq:PSFPowerLaw}
    D_\psi(r)\approx\frac{4\pi^2(3-\beta)K^2\Delta L M_\sigma^2}{\beta}r^2\times
    \begin{cases}  
    L_0^{-3}\left(\frac{L_0}{l_0}\right)^{\beta-3}& \beta < 3 \\
    -L_0^{-3}& \beta > 3 \\ 
    \end{cases}
\end{equation}
where $K=-8\pi(1+z_L)G/(\lambda c^2)$ and $M_\sigma$ is the RMS of the matter fluctuations within a cell of size $L_0^3$.

If $\beta>3$, as suggested by \cite{macquart_scattering_2004}, the mass variance is dominated by fluctuations at the outer scale, rendering a diffraction length of \citep{macquart_scattering_2004}
\begin{equation}\label{eq:rdiffPowerLawAbove3}
    r_{\text{diff}}=1.29\times10^{-11}\left(\frac{\beta-3}{\beta}\right)^{-1/2}(1+z_L)^{-1}\left(\frac{\nu}{1\text{GHz}}\right)^{-1}\left(\frac{M_\sigma}{10^{9}M_\odot}\right)^{-1}\left(\frac{\Delta L}{10\text{kpc}}\right)^{-1/2}\left(\frac{L_0}{10\text{kpc}}\right)^{3/2}\text{pc}.
\end{equation}
 We calculate $M_\sigma$ from the average density within the cell at a radius equal to the FRB's galactic impact parameter. The screen thickness ($\Delta L$) is set such that the product, $M_\sigma\Delta L/L_0$ is equal to the RMS mass along the FRB's path, $M_\sigma\Delta L/L_0\approx5.7\times10^{9} M_\odot$. Assuming again that the halo of FG\,181112 obeys a NFW profile with a scale radius $R_s=24\,\text{kpc}$ and a virial mass $10^{12}$\,M$_\odot$ \citep{prochaska_low_2019}, eq. (\ref{eq:rdiffPowerLawAbove3}) yields a diffractive scale of $1.0\times10^{-10}$pc and a scattering timescale of $t_\text{scatt}=280s$ for $\beta=3.5$ and $L_0=10$\,kpc. The scattering timescale has a shallower than linear dependence upon the index $\beta$ (note that the singularity at $\beta=3$ ) and is linearly proportional to $L_0^3$. The absence of an exponential scattering tail longer than $\sim 20\mu$s allows us to exclude, in the halo of FG\,181112, the presence of hierarchically clustered dark matter of a power law index $3<\beta<4$ with an outer scale $L_0\gtrsim 70$\;pc.\\
For a value of beta $0 < \beta < 3$ fluctuations at the outer scale no longer dominate the variance in the phase difference across the scattering screen and we must account for contributions at the inner scale. A derivation of the diffractive scale equation yields
\begin{equation}\label{eq:rdiffPowerLawBelow3}
    r_{\text{diff}}=1.29\times10^{-5}10^{-2\beta}\left(\frac{3-\beta}{\beta}\right)^{-1/2}(1+z_L)^{-1}\left(\frac{\nu}{1\,\text{GHz}}\right)^{-1}\left(\frac{M_\sigma}{10^{9}M_\odot}\right)^{-1}\left(\frac{\Delta L}{10\text{kpc}}\right)^{-1/2}\left(\frac{l_0}{\text{pc}}\right)^{\frac{\beta-3}{2}}\left(\frac{L_0}{10\text{kpc}}\right)^{\frac{6-\beta}{2}}\text{pc}
\end{equation}
This derivation is outlined by \cite{macquart_scattering_2004}, however his result is incorrect by a factor of $l_0^2/L_0^2$ (algebraic error). For the case of FG\,181112, eq. (\ref{eq:rdiffPowerLawBelow3}) yields a diffractive scale of $3.8\times10^{-9}$\,pc and a scattering timescale of $t_{\text{scatt}}=1.9$\,ms for $\beta=1.5$, $L_0=10$\,kpc and $l_0=1$\,pc. In this regime the scattering time has a steep, non-linear dependence on $\beta$, with scale dependence shifting slowly from the inner to the outer scale as $\beta$ increases from zero to three. Clearly the scattering time is degenerate with the choice of inner ($l_0$) and outer ($L_0$) scales and the index of the power law distribution ($\beta$), reducing the possible inferences which can be made regarding the hierarchically distributed dark matter in FG\,181112. However, from the observation of FRB\,181112 and eq. (\ref{eq:rdiffPowerLawBelow3}) we can form a bounding surface to constrain the possible values of $L_0$, $l_0$ and $\beta$, as given by:
\begin{equation}\label{eq:surfaceBound}
    l_0<\left(88.14\times10^{-2\beta}\left(\frac{3-\beta}{\beta}\right)^{-1/2}\left(\frac{L_0}{10\,\text{kpc}}\right)^{-\beta/2}\right)^{2/(3-\beta)}\text{pc}.
\end{equation}
Crucially, for a warm dark matter model, the inner scale is the free streaming scale, below which all structure is suppressed by the dynamics of a collisionless dark matter fluid. The free-streaming scale has been related to the particle mass of some dark matter candidates \citep{padmanabhan_theoretical_2000}, opening the door for FRBs to directly constrain particle mass in select dark matter models.

\section{Discussion}
As a consequence of the greatly improved temporal resolution of FRBs 181112 and 180924 we have been able to probe to much smaller mass scales than considered in previous treatments. Longer observation of FRBs would allow greater maximum temporal separations to be observed, extending the mass independent regime of any constraints to higher masses. Improvements to sensitivity will boost S/N and increase $y_{\text{max}}$ in the magnification-limited regime. A larger $y_{\text{max}}$ yields a larger cross section and consequently a greater observable lensing optical depth, thus providing a more sensitive probe to small scale structure.

FRBs captured at high time resolution represent an opportunity to explore fine structure of galaxy halos and clusters on unprecedented scales. We have focused here on the potential for FRBs to detect compact objects and derived simple constraints on non-baryonic dark matter models. The favoured $\Lambda$CDM cosmology is well known for its success describing the large scale structure of our Universe, but it faces a number of challenges on length scales below 1 Mpc \citep{bullock_small-scale_2017}. To meet these challenges, the substructure of dark halos must be understood, and, as shown, high time resolution FRBs provide us with the means to do so by directly constraining the inner scale of hierarchically clustered dark matter.

To exclude lensing with 95$\%$ confidence, a cumulative optical depth of 3.0 is required. From Figures \ref{fig:galOD} and \ref{fig:IGMOD}, we estimate the optical depth probed by an FRB similar to those considered here at a range of compact dark matter fractions. The cumulative optical depth probed by a set of FRBs is simply the summation of their individual optical depths as per eq. (\ref{eq:prob}). Hence, we can predict the number of FRBs that would be required to make a desired constraint. The number required varies with $f_{\text{DM}}$, the desired confidence level and the assumed distribution (e.g. in halos or distributed throughout the cosmic web). The cumulative optical depth required is non--linear with the desired level of confidence, and, hence, a lesser constraint of 80-90$\%$ would require a sample of 54-77$\%$ the size, respectively. Conversely, the cumulative optical depth required, is linear with the compact dark matter fraction, i.e. to exclude the compact dark matter fraction with the same confidence to below 0.5$f_{\text{DM}}$ requires a sample twice the size. 

In summary, recent FRBs detections, made at high time resolution, have revealed the potential of FRBs to probe dark matter within our Universe. The fact that FRBs have narrower temporal structure than previously assumed in gravitational lensing studies, allows searches for smaller lens masses than previously considered. The probability of observing halo lensing, in an FRB similar to 181112, is $\sim 0.017$ (assuming $f_{DM}\leq$0.35). To exclude $f_{DM}\geq0.35$, in galaxy halos, would require a sample of $\sim 170$ FRBs like FRB\,181112. The probability of observing lensing anywhere along the sightline, in an FRB similar to FRB\,181112 or FRB\,180924, is $\sim 0.023$ (assuming $f_{DM}\leq$0.35). This is a lower limit, in the sense that a large fraction of FRBs have dispersion measures that place them at higher redshifts than these two bursts and it ignores the possibility that the sample of already detected bursts favours lensed events through magnification bias. Thus, it is possible that a significant number of the sample of $>100$ FRBs known to date have been lensed, although the lower time resolution and lower S/N of a large fraction of these previous detections would substantially hinder the discoverability of any lensing signal. To exclude $f_{DM}\geq0.35$, in the IGM, would require detection of $\sim 130$ FRBs similar to FRB\,181112 or FRB\,180924. Finally, we conclude that when distributed as a uniform field of compact objects, the volume filling factor of dark matter in FG\,181112 is likely insufficient to contribute to the temporal scatter-broadening of FRBs on nanosecond to microsecond timescales. However, the gravitational scattering of FRBs does present a promising probe of hierarchically clustered dark matter.

\acknowledgments

J.P.M. and R.M.S. acknowledge Australian Research Council (ARC) Grant DP180100857. 
R.M.S. is the recipient of ARC Future Fellowship  FT190100155.
J.X.P. as a co-founder of the Fast and Fortunate for FRB
Follow-up team, acknowledges support from 
NSF grants AST-1911140 and AST-1910471.
A.T.D. is the recipient of an ARC Future Fellowship (FT150100415).
\vspace{5mm}
\bibliography{references}
\end{document}